\documentclass{IEEEtran}
\usepackage{subfigure}
\usepackage[cmex10]{amsmath}
\usepackage{amssymb}
\usepackage{epsfig}
\usepackage{cite}
\usepackage{graphicx}
\usepackage{color}
\usepackage{bm}
\usepackage{booktabs}
\usepackage{gensymb}
\usepackage{mathrsfs}
\usepackage{xfrac}
\usepackage{algorithm}
\usepackage{algorithmic}
\usepackage{multirow}
\usepackage{float}
\usepackage{enumerate}
\newfloat{figtab}{htb}{fgtb}
\makeatletter
 \newcommand\figcaption{\def\@captype{figure}\caption}
  \newcommand\tabcaption{\def\@captype{table}\caption}

\newlength{\figwidth}
\newcommand{\tabincell}[2]{\begin{tabular}{@{}#1@{}}#2\end{tabular}}



\begin{document}

\title{STARS for Integrated Sensing and Communications: Challenges, Solutions, \\ and Future Directions}

\author{Zheng Zhang,  Zhaolin Wang, Xidong Mu, Jian Chen, and  Yuanwei Liu
\thanks{Zheng Zhang and Jian Chen are with the School of Telecommunications Engineering, Xidian University, Xi'an 710071, China (e-mail: zzhang\_688@stu.xidian.edu.cn; jianchen@mail.xidian.edu.cn).}

\thanks{Zhaolin Wang, Xidong Mu, and Yuanwei Liu are with the School of Electronic Engineering and Computer Science, Queen Mary University of London, London E1 4NS, U.K. (e-mail: zhaolin.wang@qmul.ac.uk; xidong.mu@qmul.ac.uk; yuanwei.liu@qmul.ac.uk;).}

}

\maketitle
\begin{abstract}
 This article discusses the employment of simultaneously transmitting and reflecting surface (STARS) for integrated sensing and communication (ISAC) networks. First, two fundamental configurations of STARS-enabled ISAC systems are introduced, namely \emph{integrated full-space configuration} and \emph{separated half-space configuration}, as well as their respective advantages and common challenges are identified. To address the aforementioned challenges, a novel sensing-at-STARS design is proposed, where the sensing functionality is achieved at the STARS instead of at the base station. Such a design significantly improves the echo signal energy by eliminating undesired echo energy attenuation/leakage, in addition to establishing favorable echo propagation paths to facilitate sensing information extraction. We also present three practical implementations for sensing-at-STARS, including separated elements, mode-selection elements, and power-splitting elements. Each implementation enables flexible sensing-communication tradeoffs. Numerical results are provided to demonstrate the superiority of the proposed STARS-enabled ISAC design. Finally, we discuss several future research directions.
\end{abstract}

\IEEEpeerreviewmaketitle
\section{Introduction}\label{Section_1}
Driven by a variety of emerging applications, such as auto-driving, digital twins, extended reality (XR), and Metaverse, the next-generation wireless networks towards 2030 are seeking a unified versatile network paradigm \cite{W.Yuan_ISAC_Mag}. Fortunately, the concept of integrated sensing and communications (ISAC) provides a new perspective regarding the fusion of sensing and communications \cite{F.Liu_ISAC_JSAC,A.Liu_ISAC_Survey}. It encourages sharing the hardware platform and signal processing module among multiple functionalities, which is regarded as a fundamental shift in the wireless network paradigm. On the one hand, by designing the specialized inter-functional cooperation mechanism, ISAC can dramatically raise the utilization of the network resources, e.g., spectrum efficiency and spatial degrees of freedom (DoFs), whilst significantly reducing hardware costs. On the other hand, ISAC provides easier access to real-time channel information surveillance for communication users, which contributes to precise spatial beamforming, power allocation, and interference management. In view of the above benefits, ISAC has been deemed a key enabler for future wireless networks.

Benefiting from the materials discipline, another promising technique, namely reconfigurable intelligent surface (RIS), has been proposed to overcome the negative signal propagation issues (e.g., obstacle blockage, shadow fading, and multipath effect) in wireless networks \cite{M.Di.Renzo_RIS_JSAC}. Technically, RIS is a type of digital-domain programmable metamaterial consisting of plenty of passive reflecting elements. Each element is able to manipulate the amplitudes and/or phases of the impinging signals, thereby enabling a proactive reconfiguration of the wireless channels. More recently, it has been claimed that proper integration of RIS into ISAC networks can concurrently boost communication and sensing quality as it establishes additional line-of-sight (LoS) links for blind-zone users/targets and also mitigates the inter-functionality interference in ISAC systems \cite{X.Song_IRS_ISAC}. However, the \textit{half-space} coverage characteristics of the reflecting-only RIS restrict the sensing range and communication connectivity of ISAC systems. As a remedy, a novel concept of simultaneously transmitting and reflecting surface (STARS) is proposed \cite{Liu_magazine}. By adjusting the magnetic and electric surface reactance of the STARS, a \textit{full-space} controllable propagation environment can be provided for ISAC systems. Despite the fact that exploiting STARS exhibits new benefits for the design of ISAC systems (such as large sensing-communication (S\&C) range, extra spatial DoFs, and intuitive  S\&C tradeoff), establishing efficient STARS-enabled ISAC systems still faces tricky challenges. In particular, STARS brings a fundamentally different signal propagation, i.e., simultaneous signal transmission and reflection, which inevitably leads to the echo signal energy leakage at the STARS and introduces the inter-functionality interference between transmission and reflection regions. Hence, it calls for the redesign of the STARS architecture to unleash its potential in ISAC transmission.

Against this background, we integrate STARS into ISAC systems in this article. We commence by concisely introducing two basic configurations of STARS-enabled ISAC systems, as well as their key design challenges. To address these challenges, a new concept of sensing-at-STARS is proposed, where both operating principles and implementation methods are presented. The numerical results are provided to demonstrate the effectiveness of sensing-at-STARS structure. Finally, we make a conclusion to this article, followed by some future research directions.

\begin{figure*}[!t]
\centering
\subfigure[]{ \label{full_space}
    \includegraphics[width= 0.45\textwidth]{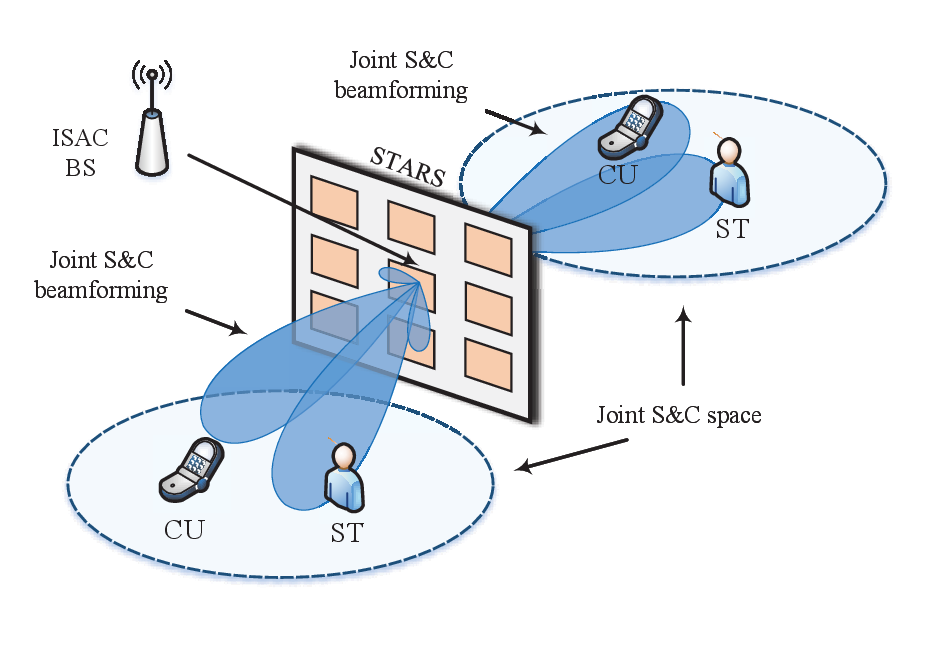}
}
\subfigure[]{\label{half_space}
    \includegraphics[width= 0.45\textwidth]{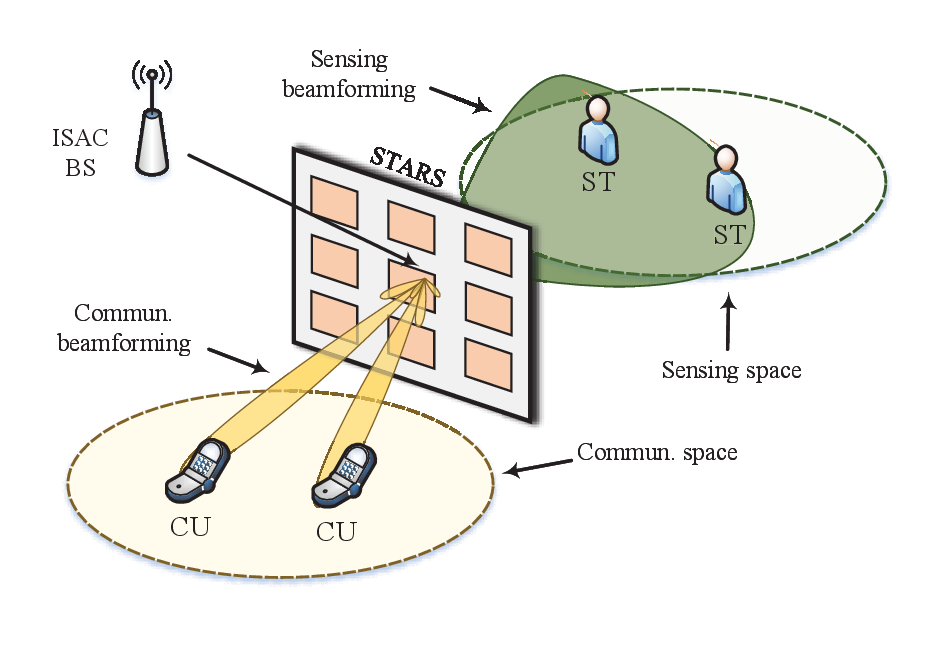}
}
\caption{Illustration of STARS-enabled ISAC systems. (a) Integrated full-space configuration. (b) Separated half-space configuration. }
\end{figure*}

\section{STARS-Enabled ISAC}\label{Section_2}

In comparison to conventional RISs, STARSs exhibit two distinctive attributes that can be harnessed for the design of ISAC systems. On the one hand, STARS enables full-space coverage. Therefore, it can facilitate seamless communication and extensive sensing across the entire space, which is referred to as \emph{integrated full-space configuration}. On the other hand, STARS also partitions the entire space into two separate spaces \cite{Z.Wang_STAR_ISAC}, namely the transmission and reflection space. This partition enables STARS to potentially accommodate communication and sensing functionalities within two separate half-spaces, which is referred to as \emph{separated half-space configuration}. In the following, we will introduce the principles, advantages, and challenges of these two configurations of STARS-enabled ISAC.

\subsection{Integrated Full-Space Configuration}
As depicted in Fig. \ref{full_space}, communication users (CUs) and sensing targets (STs) are located on both sides of the STARS in the integrated full-space configuration. The key advantages of this configuration can be summarized as follows:
\begin{itemize}
    \item \textbf{Ubiquitous S\&C Coverage:} In contrast to conventional RISs, STARS exhibits enhanced adaptability in establishing reliable LoS communication and sensing links. On the one hand, in the popular application scenarios of RISs, STARS can be more flexibly placed for establishing LoS links for randomly distributed CUs and STs, without the need for real-time orientation adjustments required by conventional RISs. On the other hand, in more stringent scenarios, such as outdoor-to-indoor and indoor, STARS can achieve S\&C coverage extension between two physically disconnected spaces by strategically deploying it on the windows and walls.
    \item \textbf{Enhanced S\&C DoFs:} With the simultaneous transmission and reflection beamforming, STARS introduces additional DoFs to favor the ISAC performance by not only directionally strengthening signals intended for CUs and STs, but also mitigating potential inter-functionality interference.
\end{itemize}

\begin{figure*}[!t]
\centering
\includegraphics[width= 0.8\textwidth]{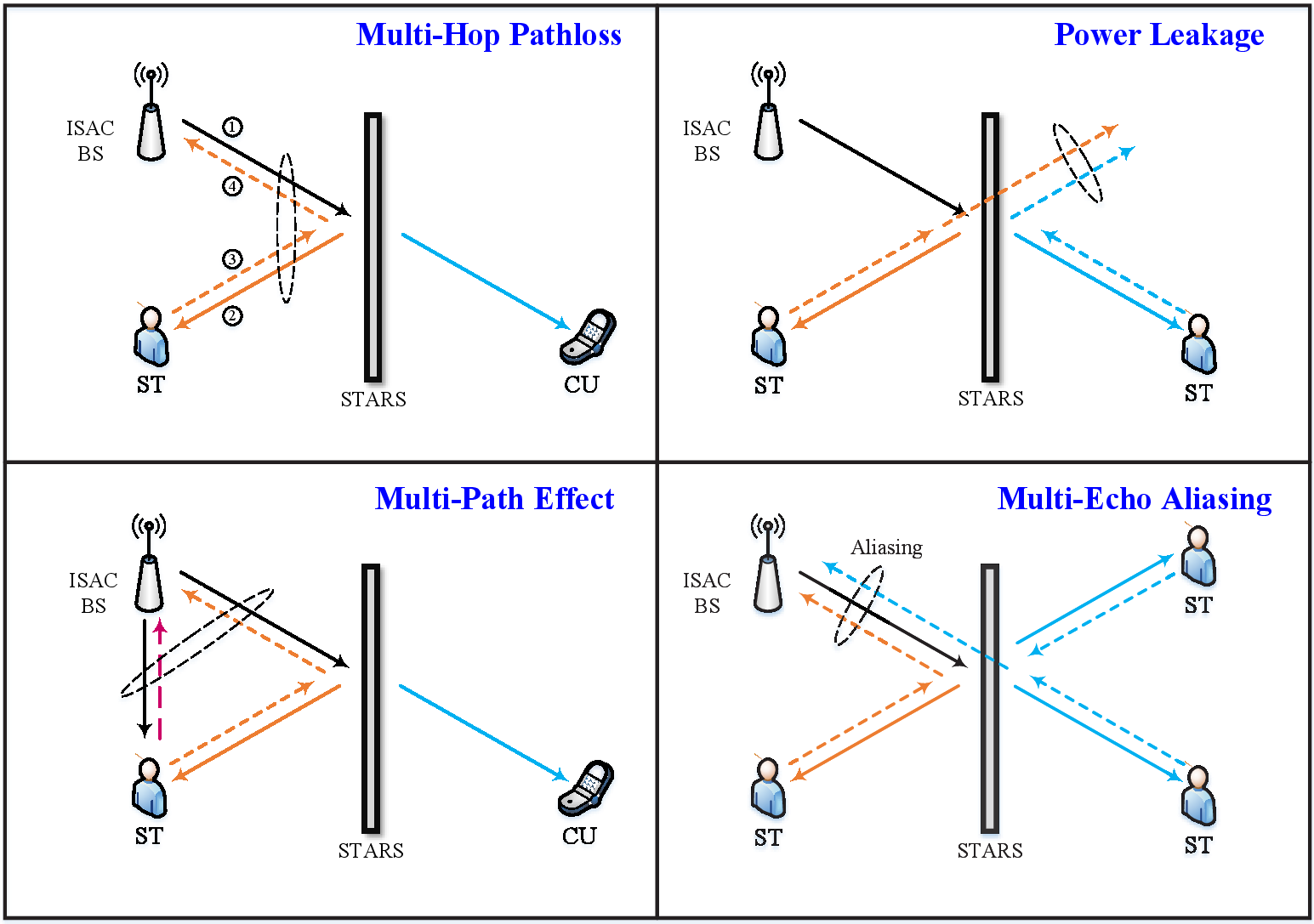}
\caption{Key design challenges for exploiting STARS in ISAC systems.}
\label{Fig2}
\end{figure*}

\subsection{Separated Half-Space Configuration}\label{Section_3}
As shown in Fig. \ref{half_space}, in the separated half-space configuration, CUs and STs are confined to distinct half-spaces on their respective sides of the STARS, thus achieving a clear demarcation between communication and sensing spaces. Although the S\&C DoFs are reduced due to the half-space configuration, it exhibits the following unique advantages:
\begin{itemize}
    \item \textbf{Independent S\&C Beamforming:} In practice, sensing and communications typically require different distinct beam configurations. For instance, the utilization of an isotropic beam proves advantageous for target detection, while directional beams are favorable for facilitating unicast communication. Fortunately, the separated half-space configuration can utilize the independent S\&C beamforming to pursue the S\&C tradeoff. In particular, the transmission and reflection signals are independently responsible for communication and sensing in the separated half-space configuration. Therefore, individualized designs for communication and sensing beamforming at the STARS become feasible, leading to a notable reduction in beamforming complexity.

    \item \textbf{Scalable S\&C Tradeoff:} In the integrated full-space configuration, the S\&C tradeoff adjustment requires the redesign of the joint beamforming. On the contrary, in the separated half-space configuration, since the S\&C beamforming can be designed independently, achieving a scalable tradeoff between S\&C becomes straightforward by adjusting transmission-reflection energy ratios, element numbers, and time allocations in energy-splitting, mode-switching, and time-sharing modes, respectively.
\end{itemize}

\subsection{Key Design Challenges}
Although both integrated full-space configuration and separated half-space configuration of STARSs have demonstrated enormous potential for S\&C performance improvement, the direct employing them in the conventional ISAC system also brings the following practical challenges (see Fig. \ref{Fig2}):
\begin{itemize}
  \item \textbf{(C1) Multi-Hop Pathloss:} Although the STARS can help to create virtual LoS links between the ISAC base station (BS) and the target, such links are subject to severe attenuation due to the multi-hop pathloss through the BS$\rightarrow$STARS$\rightarrow$target$\rightarrow$STARS$\rightarrow$BS cascaded channel. This cumulative path loss yields a considerable reduction in the power levels of the echo signals received at the BS, particularly in scenarios where direct links are obstructed. As a result, the accuracy of target detection is substantially limited. 

  \item \textbf{(C2) Echo Power Leakage:} This issue is caused by the unique dual-sided incident property of STARS. In particular, the echo signals from STs exhibit inevitable leakage towards the side of the STARS that is situated opposite the ISAC BS. Such an effect further reduces the power of echo signals received at the BS. To solve this issue, an additional ISAC BS can be deployed on the side of the STARS where leakage occurs to capture the leaked echo signals. Then, the target sensing can be carried out through cooperation between the distributed ISAC BSs.

  \item \textbf{(C3) Multi-Path Effect:} STARS results in the multi-path effect of echo signals when the direct link between BS and target exists. Conventionally, this multi-path effect can lead to the emergence of ghost targets, deceiving the ISAC BS. Although the ghost targets caused by STARS reflection and transmission can be effectively eliminated by the prior location knowledge of STARS, harnessing the potential benefits of this multi-path effect for target sensing is still challenging, which requires the high-complexity sensing receiver design \cite{ghost_target}. Moreover, in the integrated full-space configuration, the beamforming at the STARS should also be designed for communication, which aggravates the above challenge.  With the communication-prior beamforming, the indirect path caused by the STARS may be unconstructively combined with the direct path at the BS, thus resulting in reduced echo power.

  \item \textbf{(C4) Multi-Echo Aliasing:} For the multi-target scenarios, the ISAC BS receives multiple echo signals from a single STARS$\rightarrow$BS link, resulting in an aliasing effect. Such a signal aliasing renders it difficult to directly extract and process the echoes of each target and poses a high complexity to sensing algorithms.
\end{itemize}

Given the aforementioned challenges, a fundamental question arises, i.e., \emph{Is it possible to devise a unified solution that can simultaneously tackle these challenges?} The following section introduces a sensing-at-STARS approach as a response to this question, aiming to realize the desired objective.

\section{Sensing-at-STARS for ISAC}\label{Section_2}

In this section, we present the concept of sensing-at-STARS to address the above issues suffered by existing STARS-enabled ISAC networks, where three implementation methods, namely separated elements (SE), mode-selection elements (MSE), and power-splitting elements (PSE), are proposed to provide flexible full-space communication and sensing services.

\subsection{Principles and Advantages}

\begin{figure}[t!]
  \centering
  \includegraphics[width= 0.5\textwidth]{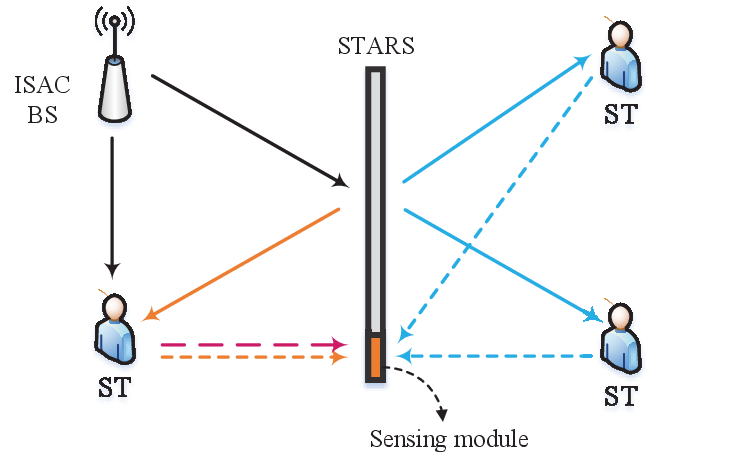}
  \caption{Illustration of the proposed sensing-at-STARS structure.}
  \label{fig_sensing_at_STARS}
\end{figure}

The core idea of sensing-at-STARS is to integrate the sensing functionality into the STARS, where the target sensing is carried out at the STARS instead of at the BS, as illustrated in Fig. \ref{fig_sensing_at_STARS}. Taking the downlink STARS-enabled ISAC transmission as an example, the BS first broadcasts the joint S\&C signals. On receiving these S\&C signals, the STARS reflects and/or transmits these signals to the STs for illumination. Afterward, the S\&C signals reflected from the targets will be received by the sensing module at the STARS, which experience a double-reflection cascaded link, i.e., BS$\rightarrow$STARS$\rightarrow$targets$\rightarrow$STARS. Finally, the subspace-based or deep learning-based sensing algorithms can be employed to extract the target information based on the received echo signals. Notably, the proposed sensing-at-STARS can efficiently address the technical challenges presented previously, which are summarized as follows:
\begin{itemize}
  \item \textbf{Low echo energy leakage:} Exploiting the sensing-at-STARS architecture renders it available to carry out sensing functionality at the STARS, which reduces the path loss that echo signals suffer from. Meanwhile, the energy leakage of the echo signals reflected/transmitted to the opposite side of the BS is fully avoided. Both of them raise the received echo energy at the sensing module, which efficiently address the challenges (C1) and (C2), thereby resulting in enhanced sensing performance.
  \item \textbf{Favorable echo propagation path:} In STARS-enabled ISAC systems with sensing-at-STARS architecture, the echo signals reflected from the single target go through the same propagation path, i.e., target$\rightarrow$sensing elements. It mitigates the multipath effects in challenge (C3). Moreover, the echo signals reflected from arbitrarily different targets experience completely non-overlapping propagation paths, which solves the echo signal overlapping issue of challenge (C4) and facilitates the detection performance for multiple targets.
\end{itemize}
Besides addressing the above challenges, exploiting sensing-at-STARS in ISAC systems also yields the following benefits:
\begin{itemize}
  \item \textbf{Higher sensing accuracy:} Applying the sensing-at-STARS architecture in existing ISAC systems, the reflected echo waves can be additionally received at the sensing module. It further increases the sampling resolution of the echo signals and improves the accuracy of the target detection.
  \item \textbf{Better adaptability:} Conventional ISAC networks require the dedicated co-design of the communication and sensing hardware or signal processing architectures. However, the sensing-at-STARS structure can be directly integrated into the pure communication networks for achieving simultaneous communication and sensing functionalities.
  \item \textbf{Low-complexity passive beamforming design:} In STARS-enabled ISAC systems without sensing-at-STARS, the sensing signals go through two signal propagation reconfigurations at the STARS, which results in a strong self-coupling of the STARS coefficients in the cascaded channels of sensing signals. It undoubtedly requires complex decoupling algorithms for the passive beamforming optimization. Whereas the sensing-at-STARS structure eliminates the secondary reflection/transmission of echo signals, and no additional decoupling operation will be needed. 
\end{itemize}

communication information bits are modulated on the waveforms of the S\&C signals and
\begin{figure*}[!t]
\centering
\includegraphics[width= 1\textwidth]{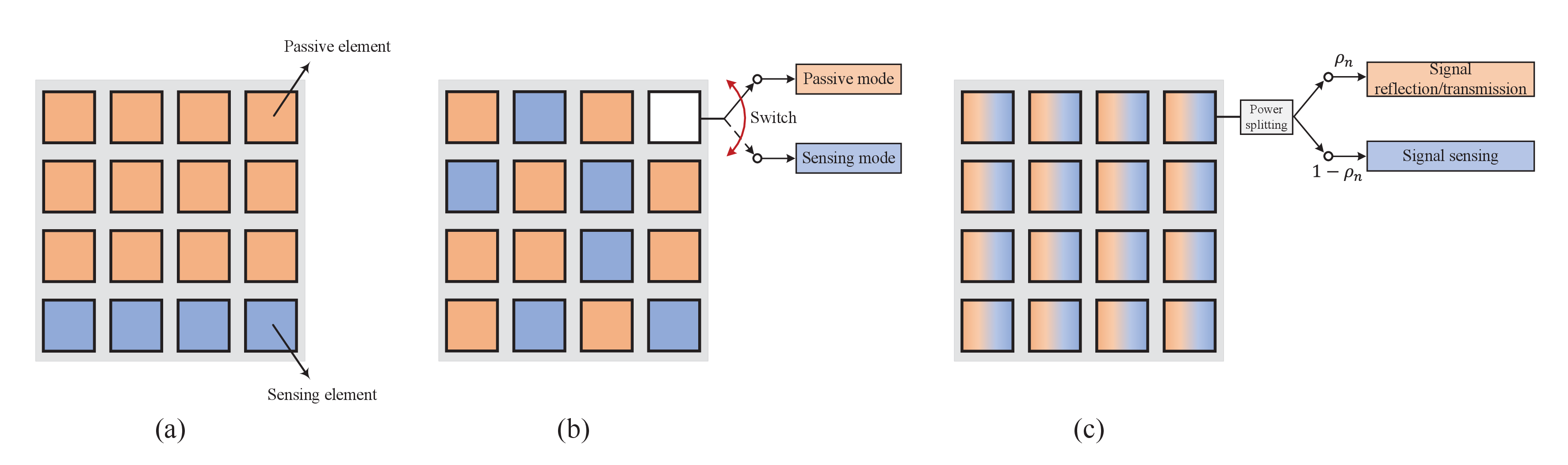}
\caption{Different implementations of sensing-at-STARS. (a) Separated elements. (b) Mode-selection elements. (c) Power-splitting elements. }
\label{fig_implementation}
\end{figure*}

\begin{table*}[t!]
  \caption{Comparison of Different Implementation Methods}
  \centering
  \begin{tabular}{|l|l|l|l|c|}
      \hline
      \textbf{Implementation Methods}      & \textbf{Advantages}       & \textbf{Disadvantages}
      \\ \hline
      Separated element                     &  Easy to implement and low hardware cost   & Low adaptability and flexibility
      \\ \hline
      Mode-selection element               & \tabincell{l}{Enabling tradeoff between number of  \\ passive-mode elements and the number of \\  sensing-mode elements.}   & Challenging element-mode optimization                                    \\ \hline
      Power-splitting element            & High echo signal sampling resolution  & Low received sensing SNR at each element                                   \\ \hline
  \end{tabular}
  \label{table1}
\end{table*}
\subsection{Implementations for sensing-at-STARS}
Toward achieving sensing-at-STARS functionality, we propose three practical implementation methods. We also identify the respective advantages and disadvantages of each implementation method.
\subsubsection{Separated Elements (SE)}
As shown in Fig. \ref{fig_implementation}(a), the STARS is equipped with two categories of basic elements, namely passive elements and sensing elements, where the functionality of each basic element is pre-determined and cannot be changed. By adjusting the electric/magnetic surface reactance or the geometrical characteristics (such as the relative distance between the substrate and the dielectric element), each passive element is capable of achieving simultaneous signal transmission and reflection, full signal reflection, and full signal transmission. For sensing elements, the active sensing module is installed inside the transparent substrates to receive and analyze the echo signals reflected from the targets. By optimizing reflection/transmission coefficients at the passive elements, the passive beam gains targeting CUs and STs can be adjusted to enhance communication and sensing performance.

From a hardware design perspective, the SE structure can be achieved by directly integrating the dedicated low-cost sensors into the STARS element. To elaborate, each sensing element encapsulates a micro-sized sensor inside the transparent substrate, which is located between two reconfigurable elements. By controlling the operation mode of the outer reconfigurable elements (e.g., full signal transmission or full signal reflection), the sensing elements can switch between bilateral sensing coverage and unilateral sensing coverage. Obviously, the SE structure is based on the existing STARS elements and the encapsulated sensors, which enables it easy to implement and enjoy low hardware cost. However, the predesigned hardware design restricts the adaptability and flexibility of the SE structure. Meanwhile, the limited number of sensing elements also limits the range and accuracy of the target sensing.

\subsubsection{Mode-Selection Elements (MSE)}
In the MSE implementation (see Fig. \ref{fig_implementation}(b)), each element can switch between two different modes, i.e., passive mode and sensing mode. Specifically, the element receives the echo signals and carries out the target detection in the sensing mode while reflecting and/or transmitting incident signals in the passive mode. By jointly optimizing element modes and passive beamforming, the MSE implementation provides a more flexible ISAC design than the SE implementation.

To realize the controllable sensing functionality, an additional nano-controller network is integrated into the reconfigurable element \cite{MS}. By monitoring the dissipated power variation of the element, the electric/magnetic surface impedance can be adjusted to a specific value corresponding to the full absorption of the echo signals, where the incidence characteristics of the echo signals can be efficiently detected. When operating in the passive mode, we switch off the nano-controller network and recalibrate the surface impedance, which turns the reconfigurable element into a passive element for signal reflection and/or transmission. Compared to the SE implementation, the MSE implementation additionally considers the optimization of the element mode selection, which results in a high ISAC performance. Nevertheless, it imposes challenges on the design of the passive beamforming of STARS.



\begin{figure*}[!t]
\centering
\subfigure[]{\label{Benchmark}
    \includegraphics[width= 0.47\textwidth]{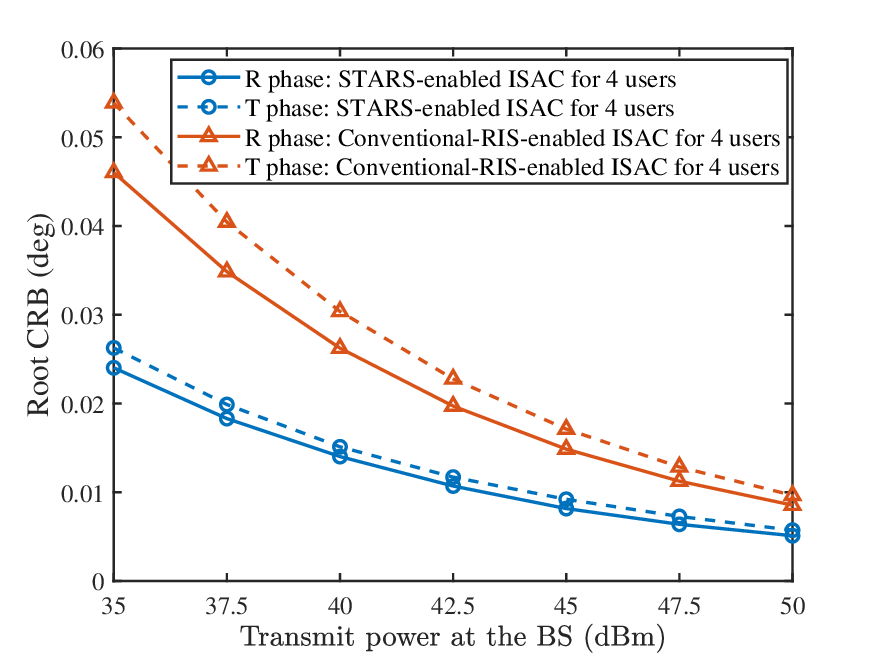}
}
\subfigure[]{\label{trade_off}
    \includegraphics[width= 0.47\textwidth]{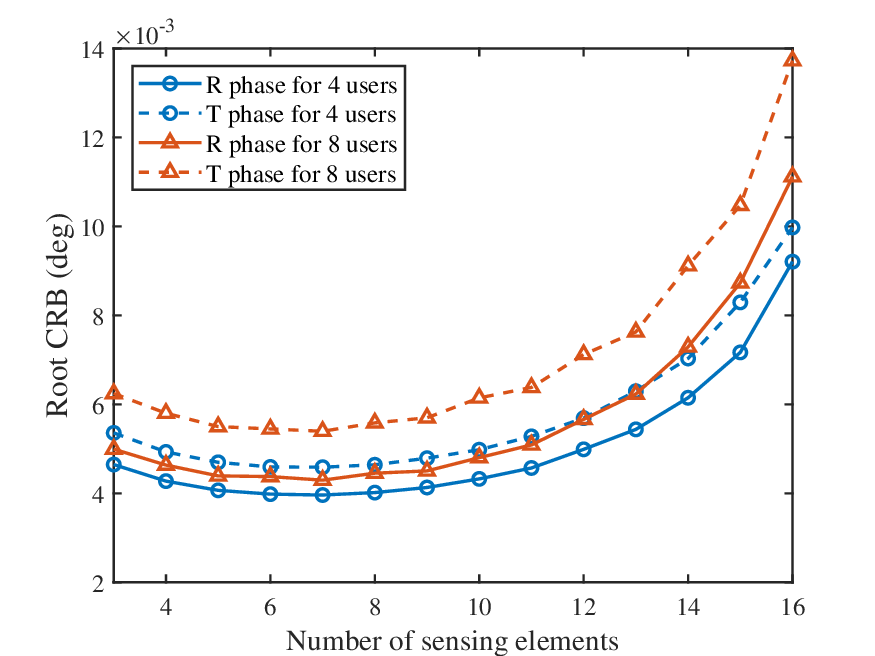}
}
\caption{Degree-based root CRB versus the transmit power at the BS and the number of sensing elements. The QoS rates for all the users are set to 0.5 bps/Hz, and other detailed simulation parameters can be found in \cite{ZZ_ISAC}: (a) performance comparison with the baseline scheme; (b) tradeoff between the number of passive elements and the number of sensing elements. }
\end{figure*}
\subsubsection{Power-Splitting Elements (PSE)}
In the PSE implementation (shown in Fig. \ref{fig_implementation}(c)), the reflection/transmission and sensing functionalities coexist in a single element. Particularly, each element can simultaneously reflect/transmit a part of incident signals while leaking the remaining part to the sensing module for parameter estimation. Obeying the law of conservation of energy, the sum energy of the sensed and reflected/transmitted signals is equal to that of the incident signals. Besides taking into account the optimization of passive beamforming, PSE implementation also requires the power splitting ratio allocation, which can further boost the S\&C performance of ISAC systems.

The PSE implementation relies on the positive-intrinsic-negative (PIN) diode-based reconfigurable element, where an annular slot is designed to couple the incident signals to the adjacent waveguide \cite{PS}. Based on this hardware implementation, a fraction of incident signals can be reflected/transmitted by the reconfigurable element, while the other fraction is captured by the waveguide connected to an RF chain. By altering the geometrical and/or spatial characteristics of the waveguide, the coupling level of the waveguide and reconfigurable element can be manipulated, so as to modify the power splitting ratio between the reflected/transmitted and sensed signals \cite{Yonina_Hybrid_RIS}. For this implementation, it becomes available to sense impinging signals at all the elements, which increases the sampling resolution of the echo signals, so as to favor the target sensing. However, since each element can only sense a fraction of the incident signals, the received sensing signal-to-noise ratio (SNR) at each element is degraded.

The comparison of three implementations is summarized in Table \ref{table1}.
In the SE implementation, the S\&C tradeoff can be achieved by the passive beamforming design at the STARS. Particularly, by adjusting the spatial directivity of the passive beamforming, it is capable of artificially controlling the received power levels of communication and sensing signals at the receiving ends, thereby realizing the dynamic tuning between the communication and sensing performance. The key tradeoff in the MSE implementation is to determine the number of passive elements and the number of sensing elements. Specifically, under the fixed total number of elements of the STARS, increasing the number of sensing elements can enhance the received energy and the sampling resolution of echo signals while deploying more passive elements can boost both the communication and sensing performance. In the PSE implementation, the elaborate power splitting ratio of each element is required to strike the S\&C tradeoff. Intuitively, allocating more power to the sensing module increases received sensing SNR and is conducive to target sensing while assigning more power to the reflected/transmitted signals can compensate for the communication signal degradation caused by the double path loss.


\section{Numerical Case Studies}
This section provides the numerical results for a study case of a STARS-enabled uplink ISAC system with sensing-at-STARS structure, where the SE implementation is considered. We consider an integrated full-space ISAC scenario, where the STARS serves to establish the reliable LoS channels from an 20-antenna BS to multiple users and two targets. All the users are randomly distributed on a circle with 20-meter (m) radius. Target 1 and target 2 are on a circle with a radius of 10 m in the directions of $(342^{\circ},30^{\circ})$ and $(18^{\circ},30^{\circ})$. To avoid the energy leakage of communication signals in uplink transmission, a STARS-enabled two-phase framework is considered. In the first phase, an 20-element STARS, with full-reflection mode passive elements, is exploited to serve the users in the reflection region while sensing target 1. In the second phase, the passive elements switch to the full-transmission mode, and the STARS aims to support the users in the transmission region while sensing target 2.

For this system setup, the joint optimization of the sensing waveform at the BS and the passive beamforming at the STARS is studied to minimize the Cram$\acute{\text{e}}$r-Rao bound (CRB) of sensing targets under the quality of service (QoS) requirements. A conventional-RIS-enabled baseline scheme is considered for comparison. To elaborate, we deploy a reflecting-only RIS and a transmitting-only RIS at the same location of STARS, each of which is equipped with $\frac{N}{2}$ elements. For fairness, we consider the sensing-at-RIS structure for the conventional RIS \cite{X.Shao_IRS_ISAC}.




Regarding the sensing performance, it is observed from Fig. \ref{Benchmark} that STARS achieves higher sensing accuracy than the conventional RIS. It validates the superiority of the STARS: 1) the deployment of the STARS increases the spatial DoFs for facilitating passive beamforming design; 2) the sensing-at-STARS structure yields more sensing elements than the sensing-at-RIS structure, which is conducive to sensing accuracy enhancement. We also observe that the sensing-at-STARS structure performs better during the reflection phase. This is expected since a portion of communication signals emitted from users are reflected back to the target, which leads to enhanced echo signal energy and a high sensing performance.

With the total number of STARS elements fixed, Fig. \ref{trade_off} reveals the tradeoff between the number of passive elements and the number of sensing elements. On the one hand, increasing the number of sensing elements benefits the echo sampling resolution enhancement, but degrades the passive beamforming gain at the STARS. On the other hand, increasing the number of passive elements introduces more DoFs to reconfigure the communication and sensing signal propagation, which however directly deteriorates the echo reception at the STARS. However, the latter plays a dominant role in STARS-enabled ISAC transmission. Moreover, it can observed that an upward trend of the number of users reduces the sensing performance of the considered system. It is because the configuration of the passive elements has to be aligned with communications users for accommodating their QoS demands, which degrades the received echo signal energy at the sensing elements.

\section{Conclusions and Future Directions}
In this article, the integration of STARS into ISAC systems was investigated. Two basic STARS-enabled ISAC configurations, namely integrated full-space configuration and separated half-space configuration, were presented. Both their exclusive benefits and design challenges were discussed. To deal with the above challenges, a novel concept of sensing-at-STARS was introduced for ISAC systems, which aimed to migrate the sensing functionality from the BS to STARS. Furthermore, three distinctive implementation methods for sensing-at-STARS were proposed to strike  the tradeoff between communications and sensing. As an emerging solution for ISAC systems, STARS also inspired some promising research directions, which are summarized as follows:
\begin{itemize}
  \item \textbf{Near-field STARS-ISAC:} To provide reliable passive beamforming gain, STARS usually needs to consist of a large-scale number of elements in practice. Such a hardware setup inevitably leads to a large STARS aperture and extends its near-field region to the ten- or even hundred-meter scale \cite{NF_mag}. Near-field signal propagation relies on the unique spherical-wave channel model. It not only introduces the extra spatial DoFs for multiplexing enhancement but also renders it possible to simultaneously estimate the distance and angle information of targets. To reap these advantages, the low-complexity passive beamforming design scheme and the tailored near-field sensing algorithms are required.

  \item \textbf{Fluid antenna for STARS-ISAC:} To overcome the severe signal attenuation caused by the obstacle block and fading effect, a novel concept, namely fluid antenna, has been proposed recently \cite{Fluid_antenna}. It can strike a good balance between the multiplexing performance and spatial diversity by adaptively adjusting the physical positions of antennas. Inspired by this, it is natural to employ the fluid antenna in the sensing-at-STARS structure. To elaborate, all the elements of STARS become separated over the whole free space, each of which can vary its own position for communication and/or sensing performance improvement. For instance, a sensing element with blocked echo links can be switched to a position with LoS echo links for sensing performance guarantee. It brings extra spatial DoFs to enhance the robustness of ISAC transmission. However, this design brings new deployment challenges to the STARS, which calls for a sophisticated element position optimization design.

  \item \textbf{NOMA for STARS ISAC:} As one of promising multiple access techniques, non-orthogonal multiple access (NOMA) permits multiple communication users share the same spectral resource with low intra-user interference. Hence, it is expected to adopt NOMA in ISAC networks to achieve flexible resource allocation and precise interference management. In details, by regarding the sensing waveforms as virtual communication signals superimposed on the real communication signals, NOMA enables receivers to employ the successive interference cancellation (SIC) to mitigate the sensing-to-communication interference in ISAC transmission. Nevertheless, the cascaded S\&C channels are highly coupled with the response coefficients of the STARS, which brings uncertainty to the SIC decoding order. As such, the dedicated passive beamforming design scheme accommodating flexible SIC decoding order is required for STARS-enabled NOMA ISAC networks.


  \item \textbf{PLS in STARS-ISAC:} The STARS-enabled ISAC system faces serious physical layer security (PLS) challenges. To elaborate, the full-space signal propagation characteristics of the STARS similarly result in a full-space eavesdropping risk. Meanwhile, integrating sensing into STARS will exacerbate this vulnerability as the information-embeded signals would be reflected/transmitted to illuminate the targets. However, these targets may possess the signal decoding ability for the purpose of maliciously eavesdropping on communicating users. Therefore, the joint S$\&$C waveform and passive beamforming design are needed to ensure the ISAC transmission while suppressing the information leakage to the targets.

\end{itemize}


\begin{thebibliography}{99}

\bibitem{W.Yuan_ISAC_Mag}
G. Caire, C. R. C. M. da Silva, T. Gu, and W. Yuan, ``Integrating sensing into communications in multi-functional networks,'' \textit{IEEE Commun Mag.}, vol. 61, no. 5, pp. 24--25, May. 2023.

\bibitem{F.Liu_ISAC_JSAC}
F. Liu, Y. Cui, C. Masouros, J. Xu, T. Han, Y. C. Eldar, and S. Buzzi, ``Integrated sensing and communications: Towards dual-functional wireless networks for 6G and beyond,'' \textit{IEEE J. Sel. Areas Commun.}, vol. 40, no. 6, pp. 1728--1767, Jun. 2022.

\bibitem{A.Liu_ISAC_Survey}
A. Liu, et al., ``A survey on fundamental limits of integrated sensing and communication,'' \textit{IEEE Commun. Surv. Tut.}, vol. 24, no. 2, pp. 994--1034, Feb. 2022.

\bibitem{M.Di.Renzo_RIS_JSAC}
M. Di Renzo, A. Zappone, M. Debbah, M.-S. Alouini, C. Yuen, J. de Rosny, and S. Tretyakov, ``Smart radio environments empowered by reconfigurable intelligent surfaces: How it works, state of research, and road ahead,'' \textit{IEEE J. Sel. Areas Commun.}, vol. 38, no. 11, pp. 2450--2525, Nov. 2020.

\bibitem{X.Song_IRS_ISAC}
X. Song, J. Xu, F. Liu, T. X. Han, and Y. C. Eldar, ``Intelligent reflecting surface enabled sensing: Cram$\acute{\text{e}}$r-Rao bound optimization,'' \textit{IEEE Trans. Signal Process.}, vol. 71, pp. 2011--2026, 2023.

\bibitem{Liu_magazine}
Y. Liu, X. Mu, J. Xu, R. Schober, Y. Hao, H. V. Poor, and L. Hanzo, ``STAR: Simultaneous transmission and reflection for 360$^{\circ}$ coverage by intelligent surfaces,'' \textit{IEEE Wireless Commun.}, vol. 28, no. 6, pp. 102--109, Dec. 2021.


\bibitem{Z.Wang_STAR_ISAC}
Z. Wang, X. Mu, and Y. Liu, ``STARS enabled integrated sensing and communications,'' \textit{IEEE Trans. Wireless Commun.}, early access, doi: 10.1109/TWC.2023.3245297.

\bibitem{ghost_target}
G. Gennarelli and F. Soldovieri, ``Multipath ghosts in radar imaging: Physical insight and mitigation strategies,'' \textit{IEEE J. Sel. Topics Appl. Earth Observ. Remote Sens.}, vol. 8, no. 3, pp. 1078--1086, Mar. 2015.

\bibitem{MS}
C. Liaskos et al., ``ABSense: Sensing electromagnetic waves on metasurfaces via ambient compilation of full absorption,'' in \textit{Proc. ACM NANOCOM}, Dublin, Ireland, Sep. 2019.

\bibitem{PS}
I. Alamzadeh et al., ``A reconfigurable intelligent surface with integrated sensing capability,'' \textit{Scientific Reports}, vol. 11, no. 1, pp. 1--10, 2021.

\bibitem{Yonina_Hybrid_RIS}
G. C. Alexandropoulos, N. Shlezinger, I. Alamzadeh, M. F. Imani, H. Zhang, and Y. C. Eldar, ``Hybrid reconfigurable intelligent metasurfaces: Enabling simultaneous tunable reflections and sensing for 6G wireless communications'' [Online]. Available: https://arxiv.org/abs/2104.04690

\bibitem{ZZ_ISAC}
Z. Zhang, Y. Liu, Z. Wang, and J. Chen, ``STARS-ISAC: How many sensors do we need?'' \textit{IEEE Trans. Wireless Commun.}, early access, doi: 10.1109/TWC.2023.3285795.

\bibitem{X.Shao_IRS_ISAC}
X. Shao, C. You, W. Ma, X. Chen, and R. Zhang, ``Target sensing with intelligent reflecting surface: Architecture and performance,'' \textit{IEEE J. Sel. Areas Commun.}, vol. 40, no. 7, pp. 2070--2084, Mar. 2022.

\bibitem{NF_mag}
H. Zhang, N. Shlezinger, F. Guidi, D. Dardari, and Y. C. Eldar, ``6G wireless communications: From far-field beam steering to near-field beam focusing,'' \textit{IEEE Commun Mag.}, vol. 61, no. 4, pp. 72--77, Apr. 2023.

\bibitem{Fluid_antenna}
K. -K. Wong, A. Shojaeifard, K. -F. Tong, and Y. Zhang, ``Fluid antenna systems,'' \textit{IEEE Trans. Wireless Commun.}, vol. 20, no. 3, pp. 1950--1962, Mar. 2021.


\end{thebibliography}
\end{document}